\documentclass[aps,floatfix,twocolumn,a4paper,showpacs, nofootinbib,superscriptaddress,10pt]{revtex4}
\usepackage[dvipsnames]{xcolor}
\usepackage{graphicx}
\usepackage{color,float,amssymb,amsmath,upgreek}
\usepackage{graphicx,float}\usepackage{graphicx,float}
\usepackage[all]{xy}
\usepackage{amsmath,upgreek}
\usepackage{amssymb}
\usepackage{color}
\usepackage{epsfig}		
\usepackage{graphicx,epstopdf}
\usepackage{subfigure}
\usepackage{pdfpages}
\usepackage[colorlinks,hyperindex]{hyperref}

\usepackage[colorlinks]{hyperref}
\usepackage{graphicx} 
\usepackage{color,float,amssymb,amsmath,upgreek}
\definecolor{green1}{RGB}{0,128,0} 
\hypersetup{hidelinks,backref=true,pagebackref=true,hyperindex=true,colorlinks=true,breaklinks=true,urlcolor= blue}
\hypersetup{%
  colorlinks = true,
  linkcolor  = blue,
  citecolor = green1,
}
\usepackage{bookmark,textgreek}
\usepackage{hyperref,color,xcolor}
\hypersetup{hidelinks,hyperindex=true,colorlinks=true,breaklinks=true,urlcolor= blue}
\hypersetup{%
  colorlinks = true,
  linkcolor  = blue
}

\begin{document}

\title{ Thermal dissociation of heavy mesons and configurational entropy    }

\author{Nelson R. F. Braga}\email{braga@if.ufrj.br}
\affiliation{Instituto de F\'{\i}sica,
Universidade Federal do Rio de Janeiro, Caixa Postal 68528, RJ
21941-972 -- Brazil}
 
\author{Luiz F.  Ferreira  }\email{luizfaulhaber@if.ufrj.br}
\affiliation{Instituto de F\'{\i}sica,
Universidade Federal do Rio de Janeiro, Caixa Postal 68528, RJ
21941-972 -- Brazil}

\author{Rold\~ao da Rocha}\email{roldao.rocha@ufabc.edu.br}
\affiliation{CMCC, Universidade Federal do ABC, UFABC, 09210-580, Santo Andr\'e, Brazil}


\begin{abstract} 
 We study the dissociation of heavy vector mesons in a thermal medium from the point of view of the configurational entropy.
 Bottomonium and charmonium vector mesons in a plasma are represented using a holographic  AdS/QCD model.  The quasinormal modes, describing the quasiparticle states, are determined for a representative range of temperature and their corresponding CE is then computed. 
 The main result indicates that the CE  is an important contrivance to investigate the stability of the heavy mesons against dissociation in the thermal medium. 
 
\end{abstract}


\maketitle


\section{Introduction}

 The configurational entropy (CE) is a quantity that implements the concept of informational entropy as an information logarithmic measure. The CE comprises the way how the information is compressed into any physical system  \cite{Gleiser:2011di,Gleiser:2012tu}. The CE represents a measure of information that regards the spatial complexity of a localized system, describing the data that is necessary to depict the spatial shape of square-integrable scalar fields \cite{Gleiser:2014ipa,Sowinski:2015cfa}. 
 The modes in a system with lower CE 
 have been shown to be more dominant, and hence more detectable and observable  \cite{Bernardini:2016hvx}.  Further aspects of the CE were presented in Ref.  \cite{Gleiser:2018kbq}. 

Regarding quantum chromodynamics (QCD), describing how gluons and quarks strongly interact, the CE has been playing a prominent role together with  holographic AdS/QCD models, to reveal important features
regarding QCD phenomenology. For example, light flavour mesonic excitations \cite{Bernardini:2016hvx} and scalar glueballs \cite{Bernardini:2016qit} were scrutinized using the CE, being further generalized in a dynamical holographic model setup \cite{Barbosa-Cendejas:2018mng}.  Besides, the CE underlying bottomonium and charmonium at zero temperature confirmed the experimental predominance of lower $S$ wave resonances and lower masses quarkonia states, in Ref.  \cite{Braga:2017fsb}. Still in the AdS/QCD paradigm, $\rho$ and $\phi$ mesons were studied  in Ref. \cite{Karapetyan:2018oye}, besides the color-glass condensate in the quark-gluon  plasma (QGP) \cite{Karapetyan:2016fai,Karapetyan:2017edu}, both  in the context of the CE. Moreover, solitonic solutions of the KdV equation, arising in a cold QGP were also studied with the instruments of the CE, in Ref. \cite{daSilva:2017jay}. 
The CE applications to heavy ion collisions (HICs) was scrutinized in Ref. \cite{Ma:2018wtw}.

The CE was shown to be, moreover, very useful 
as an apparatus to probe gravity and related phenomena. For example, 
the Hawking--Page  phenomenon of phase transition leading thermal radiation into AdS--Schwarzschild black branes was investigated  in Ref.  \cite{Braga:2016wzx} in the context of the CE. After the seminal applications of the CE to the study of stellar stability in Refs.   
 \cite{Gleiser:2013mga,Gleiser:2015rwa}, Bose--Einstein condensates of gravitons were also scrutinized, where the  critical stellar density was derived as a global minimum of the CE, in Ref.   \cite{Casadio:2016aum}. The CE was further employed to investigate domain walls  \cite{roldao} and vortexes \cite{Correa:2016pgr}, and their applications in field theory \cite{Alves:2014ksa,Alves:2017ljt}.  {\color{black}{Besides, spectral functions of bottomonium states were used in the context of the information entropy method \cite{Aarts:2014cda}.}}
 
 HICs, produced in particle accelerators, lead to the formation of a QGP. This deconfined phase of hadronic matter survives only for very short time scales, being indirectly observed,  analysing the collisions yields in the detectors. 
In particular, the fraction of heavy vector mesons detected can provide information about the plasma state. The  suppression of such particles in the by products of a HIC, as compared to a proton-proton collision,  serves as an indicator that a thermal medium was formed \cite{Matsui:1986dk,Satz:2005hx}.

An interesting tool to describe heavy vector mesons is to use the AdS/QCD holographic models. Inspired on gauge string AdS/CFT duality, these  models are particularly useful in the description of hadrons in a  thermal medium. 
A holographic model for $b \bar b$ and $ c \bar c $   vector mesons (bottomonium and charmonium respectively), involving two dimensionfull parameters,  was proposed in Refs. 
\cite{Braga:2015jca,Braga:2016wkm}. Then, an improved holographic model that contains three energy parameters  appeared in Refs. \cite{Braga:2017oqw,Braga:2017bml,Braga:2018zlu}. Two of these parameters represent  the quark mass and  the string tension, necessary in order to fit the mass spectra of quarkonium states. The third parameter represents an ultraviolet (UV) energy scale, needed to fit the spectra of decay constants. They are associated with non-hadronic decay,  proportional to the transition matrix from a meson state at excitation level $n$ to the hadronic vacuum, $  \langle 0 \vert \, J_\mu (0)  \,  \vert n \rangle = \epsilon_\mu f_n m_n $.  The UV energy parameter is related to the large mass change that occurs in such a process, when  heavy vector mesons decay to  light leptons. 
 
  At finite temperature, the mesons are described by the thermal spectral function, that composes the imaginary part of the retarded Green's function.  The relevant part of the retarded Green's function is the 2-point function, that at zero temperature  has a spectral decomposition  in terms of masses $m_n$ and   decay constants $f_n$  of the states, 
 \begin{equation}
\Uppi (p^2)  = \sum_{n=1}^\infty \, \frac{f_n^ 2}{(- p^ 2) - m_n^ 2 + i \epsilon} \,.
\label{2point}
\end{equation} 
The imaginary part of Eq. (\ref{2point}) represents a sum of Dirac delta peaks, with coefficients that are proportional to the square of the decay constants, $ f_n^2 \, \delta ( - p^2 - m_n^2 )$.  At finite temperature,  the peaks get smeared, as the temperature $T$ of the medium increases. This connection between the spectral function and the decay constants is the reason why a consistent  extension of a hadronic model to finite temperature must be consistent with decay constants.  
The results that emerge from Refs. \cite{Braga:2016wkm,Braga:2017oqw,Braga:2017bml,Braga:2018zlu} show that as the temperature increases, the heavy vector mesons dissociate in the medium. 
In this letter we want to see how this dissociation process is translated into the CE. 

This paper is organized as follows: Sect. II is devoted to briefly introducing the motivation and the key points in the CE paradigm. The 2-point correlators and their power spectrum are shown to give rise to a correlation probability distribution, the modal fraction, that is the leading ingredient to compute the CE. In Sect. III the holographic model is introduced for heavy vector mesons, whose pure gauge action yields normalizable solutions, used for deriving the decay constants for charmonium and bottomonium states, whose quasinormal modes in the finite temperature case are derived  and analyzed. Sect. IV is then dedicated to the compute the CE underlying quarkonia, at finite temperature, as a function of the $S$-wave resonances excitation level and the temperature as well. Relevant results impact the  probability of dissociation with respect to the temperature, distinguishing the charmonium from the bottomonium. 

\vspace*{-0.3cm}
\section{Configurational entropy and Shannon information entropy}
 Given a probability distribution, the CE, motivated by the Shannon's  theory of information, is defined for discrete systems by $S[f] = -\sum_a p_a\,\log(p_a)$, standing for an upper limit of the information compression \cite{Gleiser:2012tu,Sowinski:2015cfa}. The CE, thus, comprises  
the informational bulk of all configurations in a physical system.  If all the $N$ system modes present an uniform 
probability distribution, then one can assume $p_a = 1/N$, yielding  the CE to attain its maximum value $S = \log N$. 
The (differential) CE takes into account the continuous limit of 
this paradigm \cite{Gleiser:2018kbq}, whose 
equivalent features were also discussed in Ref. \cite{Bernardini:2016hvx}. 

To introduce the CE of a physical system, spatial correlations must be taken into account, characterizing the correlation of the fluctuations of
a localized, square integrable, $\rho(x)$  scalar field, that describes the system. Small perturbations can make the scalar field to fluctuate.  
The correlator, namely, the 2-point correlation function reads $G(r) = { \int_{\mathbb{R}^d}\rho(\tilde{r})\rho(r+\tilde{r})d^d\tilde{r}}$, where $d$ denotes the spatial dimension. The  CE measures the informational theory
of correlations, when constructing a probability distribution from the correlator. For it, the power spectrum $P(k)$ is taken into account as the correlator Fourier transform.  The convolution theorem makes it possible to express the power
spectrum as
$P(k) \sim {\| \int_{\mathbb{R}^d}\rho(r)e^{ik\cdot r}\,d^dr\|^2}$, being normalizable. It corresponds to the uniform distribution, if $\rho(r)$ is a random distribution.  Since the normalizable power spectrum satisfies the condition to be a continuous
probability distribution associated to some continuous random variable, taking the Fourier transform 
$\rho(k) = \int_{\mathbb{R}^d}\rho(r)e^{-ik\cdot r}\,d^dx,$ 
the correlation probability distribution, known as the modal fraction, is defined as  
\cite{Gleiser:2012tu}
\begin{eqnarray}
\upepsilon(k) = \frac{|\rho(k)|^{2}}{ \int_{\mathbb{R}^d}  |\rho(k)|^{2}d^dk}.\label{modalf}
\end{eqnarray} The modal fraction measures the contribution of a given mode, $k$, to the power
spectrum.
As the power spectrum describes the
fluctuation of the scalar field spatial profile, the modal fraction measures the contribution of a range of modes for the description of the shape of the scalar field. The amount of information to describe the spatial profile of $\rho$, in terms of the Fourier modes, is given by the CE, 
\begin{eqnarray}
S[\rho] = - \int_{\mathbb{R}^d}{\upepsilon_\circ}(k)\log {\upepsilon_\circ}(k)\, d^dk\,,
\label{confige}
\end{eqnarray}
where $\upepsilon_\circ(k)=\upepsilon(k)/\upepsilon_{\rm max}(k)$, being the denominator the maximum fraction. In the case where  the modal fraction represents the uniform distribution, it maximizes the CE, corresponding to a random field for which the power spectrum is the uniform distribution. 
 
\vspace*{-0.5cm}
\section{Holographic Model}
\label{holog}  
For the zero temperature case, the model proposed in Refs. \cite{Braga:2017bml,Braga:2018zlu}  for  charmonium and bottomonium, representing quarkonia states, is defined in a  5D anti-de Sitter (AdS) space-time 
\begin{equation}
 ds^2 \,\,= \,\, \frac{R^2}{z^2 }(-dt^2 + d{x}^i d{x}_i + dz^2)\,.
\end{equation}
The heavy vector mesons are described by a vector  field $V_m = (V_\mu,V_z)\,$ ($\mu = 0,1,2,3$), assumed to be dual to the fermionic gauge theory current density $ J^\mu = \bar{\psi}\gamma^\mu \psi \,$. The action for this field reads 
\begin{equation}
I \,=\, \int d^4x dz \, \sqrt{-g} \,\, e^{- \phi (z)  } \, \left(  - \frac{1}{4 g_5^2} F^{mn} F_{mn}
\,  \right) \,\,, 
\label{vectorfieldaction}
\end{equation}
where $F_{mn} = \partial_m V_n - \partial_n V_m$ and $\phi(z)$ is a phenomenological  background field with the form
\begin{equation}
\phi(z)=k^2z^2+Mz+\tanh\left(\frac{1}{Mz}-\frac{k}{ \sqrt{\Gamma}}\right) \,.
\label{dilatonmod1}
\end{equation} 
{\color{black}{It is important to note that this is a phenomenological effective model for heavy vector mesons.
The background that we use is not a solution of the Einstein's equations, since we included the scalar field of Eq. (\ref {dilatonmod1}). This type of model is usually called a bottom up holographic model, as opposed to the so called top down models that use backgrounds that are solutions of Einstein's equations.}}
The parameters play the following role:  $k$ represents the  quark mass, $\Gamma $ the string tension of the strong quark anti-quark interaction and $M$ is a mass scale associated with non hadronic decay. 

One chooses the gauge $V_z=0$. Then,  the equation of motion for the spatial transverse components  of the  field, denoted generically as $V$, takes in momentum space the following form, 
\begin{equation}
\partial_{z} \left[ e^{-B(z)} \partial_{z} V \right]-p^2 e^{-B(z)}V=0, 
\label{eqmotion}
\end{equation}
where   $B(z)=\log\left(\frac{z}{R}\right)+\phi(z)$.

The meson states at zero temperature are represented by the normalizable solutions, $ V(p,z)=\Psi_n(z)$, that satisfy the boundary conditions $ \Psi_n(0)=0$, for $p^2=-m_{n}^{2}$, where $m_n$ denotes the masses of the corresponding meson states.  These eigenfunctions are  normalized as $
\int^{\infty}_{0} e^{-B(z)} \ \Psi_n(z)\Psi_m(z)\,dz=\delta_{mn} $, and the decay constants are  calculated holographically as
\begin{equation}
f_n=\frac{1}{g_{5} m_{n}}\lim\limits_{z \rightarrow 0} \left( e^{-B(z)}\partial_z \Psi_n(z)\right) \,.
\label{decayconstant}
\end{equation}

 The  values of the parameters in the background (\ref{dilatonmod1}), obtained in Ref. \cite{Braga:2018zlu} 
  in order to describe  charmonium and bottomonium, are 
 respectively:
\begin{eqnarray}
  \!\!\!\!\!\!\!\!\!\!\!\!k_c &=&\! 1.2  \, {\rm GeV }, \,\,   \sqrt{\Gamma_c } = 0.55  \, {\rm GeV }, \,\, M_c=2.2  \, {\rm GeV }\,,
  \label{parameters1}\\
 \!\!\!\!\!\! \!\!\!\!\!\!k_b &=&\! 2.45  \, {\rm GeV }, \,\,   \sqrt{\Gamma_b } = 1.55  \, {\rm GeV }, \,\, M_b=6.2  \, {\rm GeV }\,.
  \label{parameters2}
  \end{eqnarray}  
  Tables I and II show the results for charmonium and bottomonium, respectively. For comparison, the experimental data from Ref. \cite{Agashe:2014kda} is show between parentheses.  It is worth to mention that the higher the radial excitation level, the lower the decay constants are. 
 \begin{table}[h]
\centering
\begin{tabular}[c]{|c||c||c|}
\hline 
\multicolumn{3}{|c|}{  Holographic (and experimental)  results for charmonium   } \\
\hline
 State &  Mass (MeV)     &   Decay constants (MeV) \\
\hline
$\,\,\,\, 1S \,\,\,\,$ & $ 2943 \,\, (3096.916\pm 0.011)  $  & $ 399 \, (416 \pm 5.3)$ \\
\hline
$\,\,\,\, 2S \,\,\,\,$ & $  3959 \,\, (3686.109 \pm 0.012) $   & $ 255  \, (296.1 \pm 2.5)$  \\
\hline 
$\,\,\,\,3S \,\,\,\,$ & $  4757 \,\, (4039 \pm 1 ) $   & $198 \, ( 187.1  \pm 7.6) $ \\ 
\hline
$ \,\,\,\, 4S  \,\,\,\,$ & $ 5426\,\,  (4421 \pm 4)  $  & $ 169 \,  (160.8  \pm 9.7)$ \\
\hline
\end{tabular}   
\caption{Holographic masses and decay constants for the charmonium $S$-wave resonances. Experimental values between parentheses for comparison.  }
\label{table1}
\end{table}
\begin{table}[h]
\centering
\begin{tabular}[c]{|c||c||c|}
\hline 
\multicolumn{3}{|c|}{  Holographic (and experimental) results for bottomonium   } \\
\hline
 State &  Mass (MeV)     &   Decay constants (MeV) \\
\hline
$\,\,\,\, 1S \,\,\,\,$ & $ 6905 \,\,(9460.30\pm 0.26) $  & $ 719 \,( 715.0 \pm 2.4) $ \\
\hline
$\,\,\,\, 2S \,\,\,\,$ & $   8871 \,( 10023.26 \pm 0.32) $   & $ 521 \,(497.4 \pm 2.2) $  \\
\hline 
$\,\,\,\,3S \,\,\,\,$ & $  10442 \, \,( 10355.2 \pm 0.5) $   & $427 \, (430.1  \pm 1.9) $ \\ 
\hline
$ \,\,\,\, 4S  \,\,\,\,$ & $ 11772 \, (10579.4 \pm 1.2)  $  & $ 375 \,(340.7  \pm 9.1)$ \\
\hline
\end{tabular}   
\caption{Holographic masses and decay constants for the bottomonium $S$-wave resonances. Experimental values between parentheses for comparison. }
\label{table2}
\end{table}
  
 For the finite temperature case, we consider the same action of Eq. (\ref{vectorfieldaction}), now in an AdS  black hole  space-time 
\begin{equation}
 ds^2 \,\,= \,\, \frac{R^2}{z^2}  \,  \Big(  -  f(z) dt^2 + \frac{dz^2}{f(z) }  + dx^i dx_i  \Big)   \,,
 \label{metric2}
\end{equation}
where
\begin{equation}
f (z) = 1 - \frac{z^ 4}{z_h^4}.
\end{equation}
 The horizon position  $z_h$ is determined from the condition $f(z_h)=0$ and the black hole temperature comes from the requirement of absence of a conical singularity at the horizon, in the Euclidean version of the metric. This condition implies a relation between the periodic time interval $ 0 \le t \le \beta = 1/T$ 
 and the horizon position, given by 
\begin{equation} 
T =  \frac{\vert  f'(z)\vert_{(z=z_h)}}{4 \pi  } = \frac{1}{\pi z_h}\,.
\label{temp}
\end{equation}

\begin{figure}[H]
\begin{center}
\includegraphics[scale=0.41]{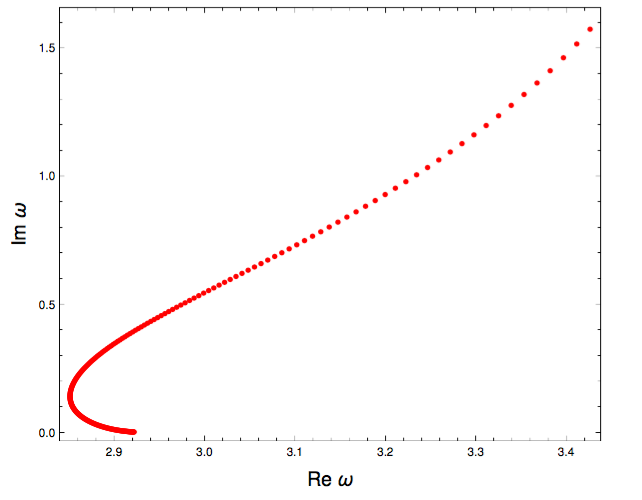}
\caption{Quasinormal modes for charmonium.}
\label{g40}
\end{center}
\end{figure}
\begin{figure}[H]
\begin{center}
\includegraphics[scale=0.36]{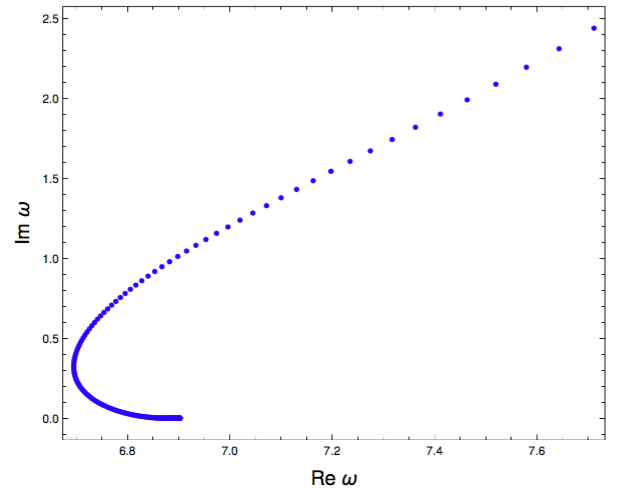}
\caption{Quasinormal modes for  bottomonium.}
\label{g41}
\end{center}
\end{figure}

In contrast to the zero temperature case, where the particles are described by normalized solutions of the field, at finite temperature the quasiparticles are described by solutions known as  quasinormal modes.   Previous studies of quasinormal modes in the context of gauge/gravity duality
can be found, for example, in Refs. \cite{Nunez:2003eq,Horowitz:1999jd,Kovtun:2005ev,Miranda:2005qx,Hoyos:2006gb,Berti:2009kk,Morgan:2009vg,Miranda:2009uw,Konoplya:2011qq,Mamani:2013ssa,Mamani:2018qzl}.
Quasinormal modes are obtained solving  the equation that comes from the metric (\ref{metric2}):
\begin{equation}\label{eqmotionT}
\partial_{z}(f(z)e^{-B(z)}\partial_{z}V)+\frac{\omega^{2}}{f(z)}e^{-B(z)}V=0\,,
\end{equation}
with the following boundary conditions:
\begin{eqnarray}\label{qsn}
v_{n}(0)&=&0 \,,\\ \nonumber
\lim_{z\rightarrow z_h}v_{n}&=& \left(1-\frac{z}{z_h}\right)^{-i\omega/(4 \pi T)}\,\left[1+a_{1}\left(1-\frac{z}{z_h}\right)+\cdots\right].
\end{eqnarray}
The second condition corresponds to selecting just infalling boundary conditions at the horizon position. The coefficients that appear in the infalling condition can be determined inserting this condition into Eq.  (\ref{eqmotionT}). Similarly to normal modes, the quasinormal modes exist only for a discrete set of frequencies $\omega_{n}$. However, in the case of quasinormal modes, these frequencies are complex.

\begin{figure}[h]
\begin{center}
\includegraphics[scale=0.35]{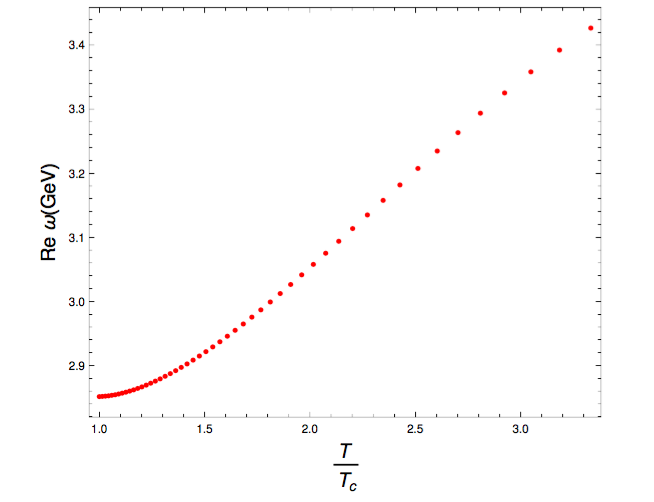}
\end{center}
\caption{ Real  part of the frequency for charmonium quasinormal modes.}
\label{g50}
\end{figure}

\begin{figure}[h]
\begin{center}
\includegraphics[scale=0.35]{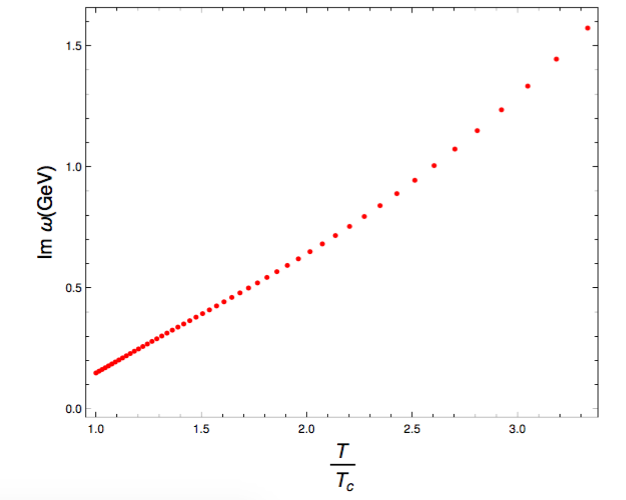}
\end{center}
\caption{Imaginary part of the frequency for charmonium quasinormal modes.}
\label{g51}
\end{figure}

\begin{figure}[h]
\begin{center}
\includegraphics[scale=0.36]{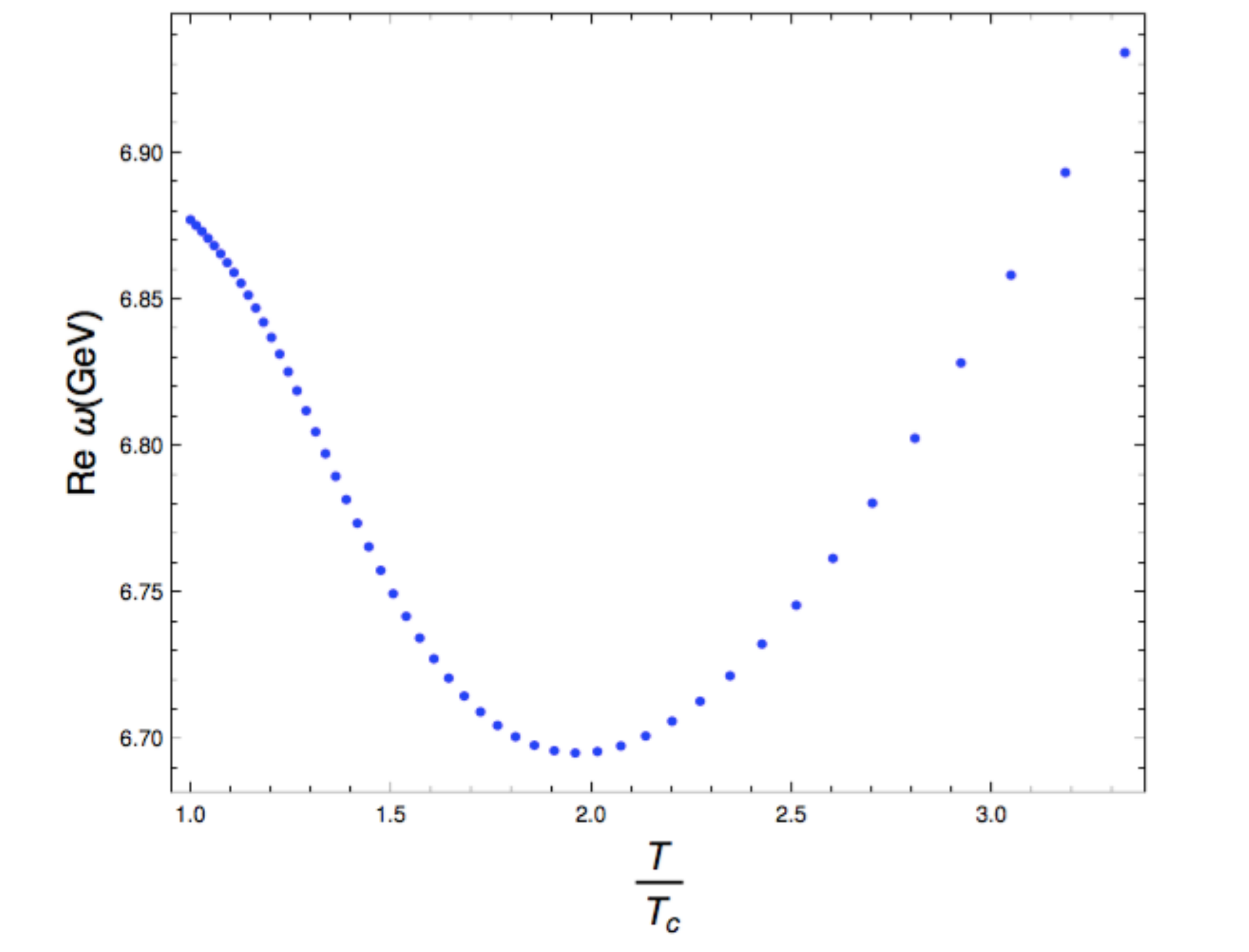}\end{center}
\caption{ Real  part of the frequency for bottomonium quasinormal modes.}
\label{g70}
\end{figure}

\begin{figure}[h]
\begin{center}
\includegraphics[scale=0.36]{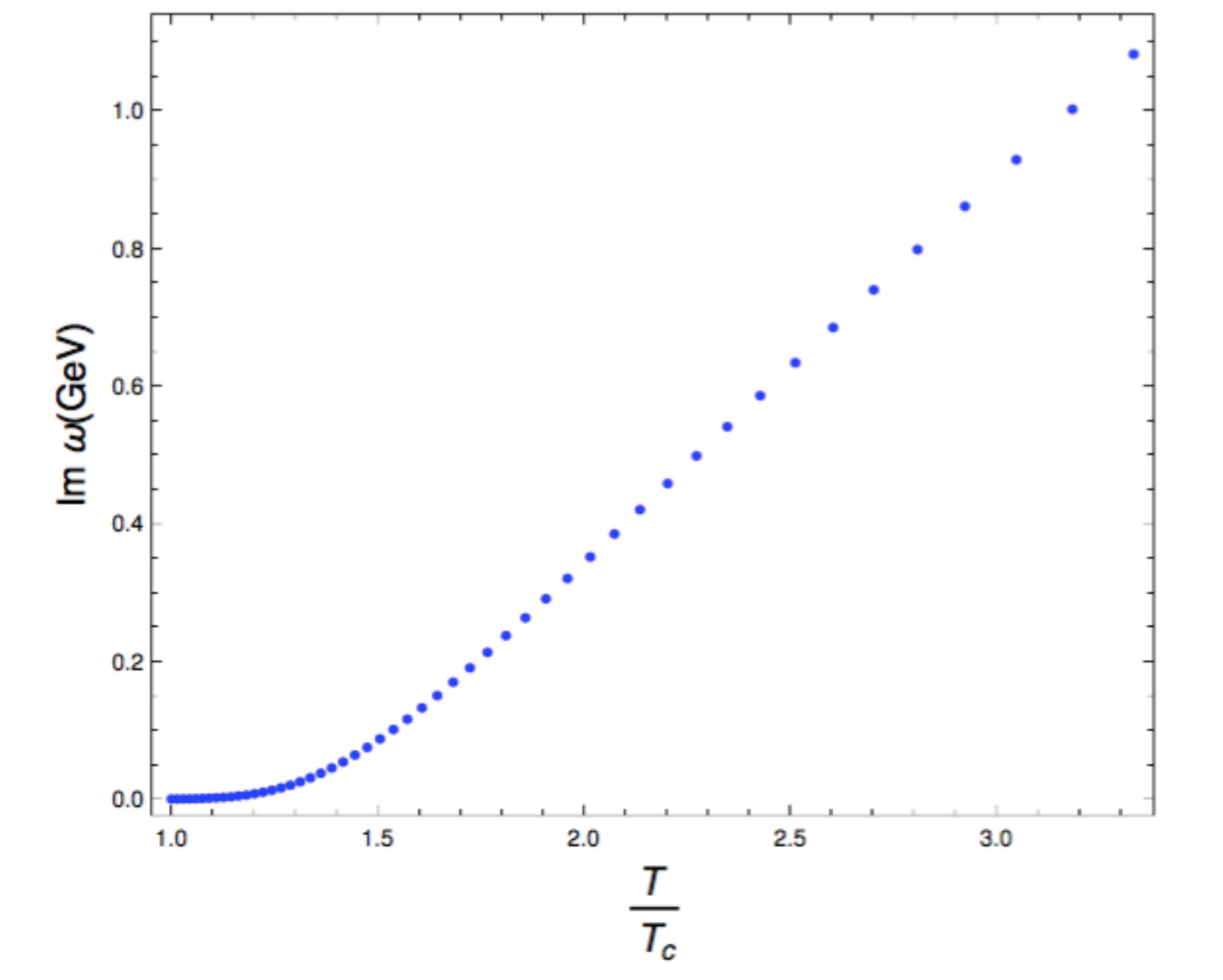}
\end{center}
\caption{ Imaginary part of the frequency for bottomonium quasinormal modes.}
\label{g71}
\end{figure}

The method used  to find the quasinormal modes in this work was the shooting method \cite{Kaminski:2008ai,Kaminski:2009ce,Janiszewski:2015ura}, whose  basic idea consists of specifying two boundary conditions at the horizon, and then adjust the free parameter given by the frequency $\omega$, to find the desired solution at the boundary. One of these two boundary conditions is the infalling boundary condition,  presented in Eq. (\ref{qsn}). The second boundary condition is the derivative of the  infalling boundary condition. It is necessary to ensure that the solution found by solving Eq. (\ref{eqmotionT}), using the boundary conditions at the horizon, is a quasinormal mode varying  the frequency $\omega$ until the field  vanishes  at the boundary.

The quasinormal modes of charmonium and bottomonium are determined as function of the temperature. We show in Figs. \ref{g40} and  \ref{g41} the results for the  imaginary part of the frequencies as a function of the real part for the first mode $n=0$. In Figs. \ref{g50} --  \ref{g71}, the real and imaginary parts of the quasinormal modes are plotted  for both the charmonium  and bottomonium respectively,  as functions of the temperature. One notes that in the zero temperature limit, the real part of the quasinormal modes coincides with the corresponding quarkonium mass.

\section{Configurational entropy of quarkonia at finite temperature}
\label{sectce}

We now apply the CE paradigm to the analysis of heavy mesons. {\color{black}{The action integral of the  holographic model that we use is not consistent with the Einstein's equations, since we include the phenomenological scalar background of Eq.}} (\ref{dilatonmod1}). {\color{black}{However, we will assume that the energy momentum tensor of the model is obtained from the action in the same way as  in general relativity. }}
 For an action integral of the form $   \int d^4x dz \, \sqrt{-g} {\mathcal L} $ the energy momentum tensor reads
\begin{equation}\label{EnTe}
T_{mn}(z)=\frac{2}{\sqrt{-g}}\!\left[ \frac{\partial(\sqrt{-g}\mathcal{L})}{\partial g^{mn}}-\frac{\partial}{\partial x^{p}}\frac{\partial(\sqrt{-g}\mathcal{L})}{\partial\left(\frac{\partial g^{mn}}{\partial x^{p}} \right)} \right]
\end{equation}
Note that in the case of the action (\ref{vectorfieldaction}), the second term of (\ref{EnTe}) is absent, since the action does depend on derivatives of the metric. The energy density $\rho (z)=T_{00}(z)$ for the case of a vector field reads
\begin{equation}
\!\!\!\!\!\!\rho (z)\!=\!\frac{e^{- \phi (z)  } }{g^{2}_{5}}\!\left[g_{00}\!\left(\frac{1}{4}g^{mp}g^{nq}F_{mn}F_{pq}\right)\!-\!g^{mn}F_{0n}F_{0m}\right]\,.
\end{equation}
Considering the metric (\ref{metric2})  and 
taking  a plane wave solution in the $x^\mu$ directions,  in the meson rest frame $V_{\mu}=\epsilon_{\mu}v(p,z)e^{-i\omega t}$, with $\epsilon_{\mu}=(0,1,0,0)$, the energy density takes the form 
\begin{equation}\label{rho}
\rho (z)=\frac{e^{-\phi(z)}}{2g^{2}_{5}R^2}\left[ -
\left(1-\frac{z^4}{z_{h}^{4}}\right)^{2}(\partial_{z}v)^{2}+\vert \omega\vert^2v^2 \right]
\end{equation}
Note that for complex solutions, the field strength term of the Lagrangian density takes the form $F^{*}_{mn} F^{ mn}$.   In Figs. \ref{g40}  --  \ref{g71}, we present the real and imaginary parts of the frequency as a function of the temperature. The solutions for $v(p,z)$ at finite temperature  are the quasinormal modes $v_n (z)$,  described in the previous section.  In the zero temperature limit, these solutions become the eigenfunctions  of Eq. (\ref{eqmotion}). Taking a Fourier transform in coordinate $z$ and splitting $\rho (k) = C(k) + iS(k)$, where
\begin{eqnarray}
C(k)&=&\int_{0}^{\infty}\rho(z)\cos({kz})dz \,,\\
S(k)&=&\int_{0}^{\infty}\rho(z)\sin({kz})dz \,,
\end{eqnarray}
one finds the  modal fraction (\ref{modalf})
\begin{equation}
{\upepsilon}(k)=\frac{S^2(k)+C^2(k)}{\int^{\infty}_{0}\left[S^2(\mathring{k})+C^2(\mathring{k})\right]d\mathring{k}}\,.
\end{equation}
Finally, the CE (\ref{confige}) reads 
\begin{equation}\label{CE}
S[\rho]=-\int^{\infty}_{0}{\upepsilon_\circ}(k)\log\upepsilon_\circ(k)\,dk\,.
\end{equation}

\begin{table}[h]
\centering
\begin{tabular}[c]{||c||c||c||}
\hline
\hline  
\multicolumn{3}{||c||}{Bottomonium} \\
\hline
\hline 
$S$-wave resonance & Masses (GeV)    &   CE \\
\hline
$\,\,\,\, 1S \,\,\,\,$ & $ 6.9051$  & $2.1786$ \\
\hline
$\,\,\,\, 2S \,\,\,\,$ & 8.8710 & 2.4957 \\
\hline 
$\,\,\,\, 3S \,\,\,\,$ & 10.442 & 2.6899 \\ 
\hline 
$\,\,\,\, 4S \,\,\,\,$ & 11.772 & 2.8188 \\ 
\hline 
$\,\,\,\, 5S \,\,\,\,$ & 12.943 & 2.9185 \\ 
\hline 
$\,\,\,\, 6S \,\,\,\,$ & 14.001 & 3.0008 \\ 
\hline 
$\,\,\,\, 7S \,\,\,\,$ & $14.972$ & $3.0712$ \\ 
\hline
\hline
\end{tabular}   
\caption{Configurational entropy (CE) of bottomonium with respect to the $S$-wave resonances excitation level.}
\label{table3}
\end{table}

\begin{figure}[h]
\!\!\!\!\includegraphics[scale=0.75]{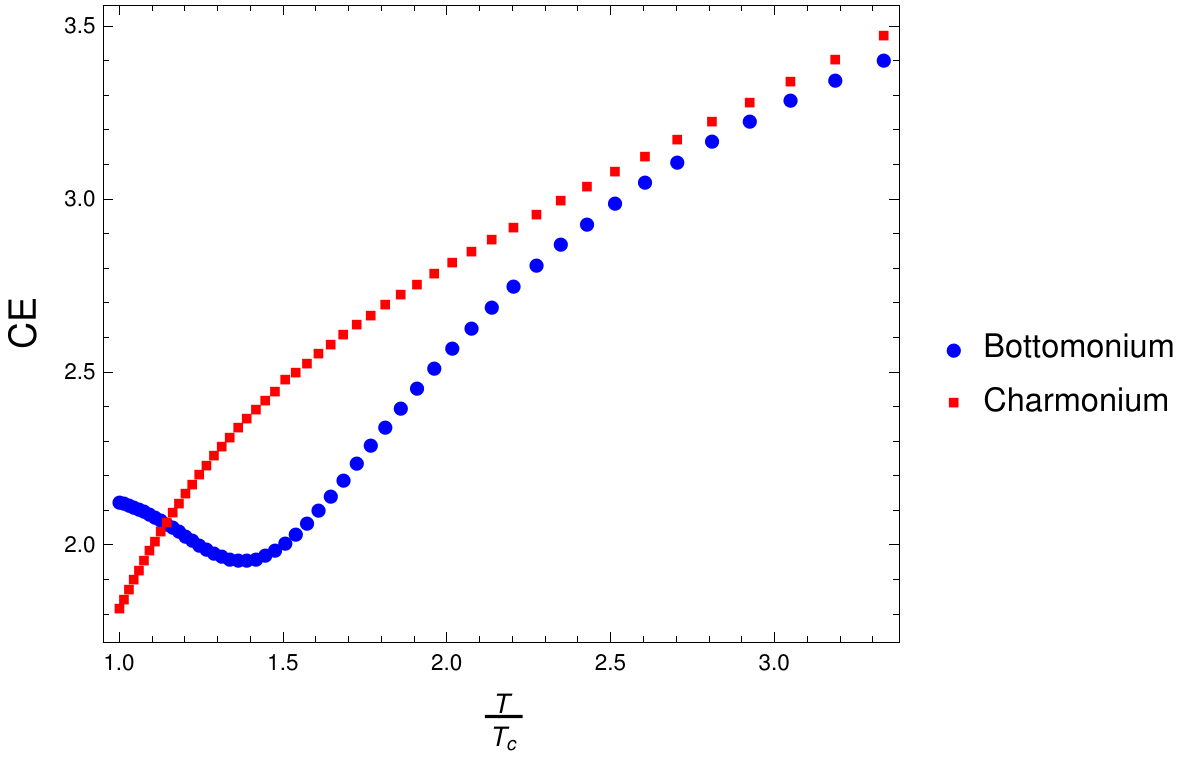}
\caption{ Configurational entropy (CE)  for bottomonium and charmonium 1$S$ quasi-state as a function of the ratio $T/T_c$.}
\label{g67}
\end{figure}

Using the appropriate values of the parameters for the quarkonia  (\ref{parameters1}) - (\ref{parameters2}) and employing Eqs. (\ref{confige}) and \eqref{rho}, we compute the CE for the quasi-states. First one evaluates  the CE for the states at  zero temperature using the eigenfunctions of  Eq. (\ref{eqmotion}). The results are listed in  Tables \ref{table3} and \ref{table4}. Then, the results for the 1$S$ quarkonia quasi-states, at finite temperature, are  presented in Fig. \ref{g67} as a function of $T/T_c$, where $T_c$ is the critical deconfinement temperature at which  the deconfined medium is formed. {\color{black}{Using lattice QCD studies, Refs. 
\cite{Aarts:2007pk,Asakawa:2003re,Fujita:2009wc,Fujita:2009ca} show that heavy mesons only melt at higher temperatures than $T_c$.}} 

 \begin{table}[h]
\centering
\begin{tabular}[c]{||c||c||c||}
\hline 
\hline 
\multicolumn{3}{||c||}{Charmonium} \\
\hline
\hline 
$S$-wave resonance & Masses (GeV)    &   CE \\
\hline
$\,\,\,\, 1S \,\,\,\,$ & 2.9425 & 1.4105 \\
\hline
$\,\,\,\, 2S \,\,\,\,$ & 3.9586 & 1.7413 \\
\hline 
$\,\,\,\, 3S \,\,\,\,$ & 4.7573 &  1.9457 \\ 
\hline 
$\,\,\,\, 4S \,\,\,\,$ & 5.4261 & 2.0793 \\ 
\hline 
$\,\,\,\, 5S \,\,\,\,$ & 6.0114 & 2.1791 \\ 
\hline 
$\,\,\,\, 6S \,\,\,\,$ & 7.0201 & 2.3130 \\ 
\hline 
$\,\,\,\, 7S \,\,\,\,$ & $7.4675$ & $2.4526$ \\ 
\hline
\hline
\end{tabular}   
\caption{Configurational entropy (CE) with respect to the $S$-wave resonances excitation level.}
\label{table4}
\end{table}

{\color{black}{From Fig. \ref{g67}, one notes that,  for the charmonium states,  the CE monotonically increases  with the temperature.  An increment in the CE is expected to represent a decrease in the stability of the system.  Charmonium quasi-states were studied in Ref. \cite{Braga:2018zlu} through the determination of the spectral function. The 1$S$ quasi-state appears as a peak that  gets continuously smeared as the temperature increases. The higher the temperature, the lower and broader is the peak,  corresponding to a continuous increment in the  degree of dissociation in the medium. The quasi-normal modes obtained in Sect. \ref{holog} show an equivalent behaviour. The imaginary part of the quasi normal mode frequency increases with the temperature. This imaginary component is related to the width of the peak. So, the continuous increament in this quantity represents the increase of the dissociation degree. Charmonium states survive the deconfinement transition, in contrast to  light hadrons that completely dissociate. But, as seen from the spectral function and from the quasinormal modes, they undergo a partial dissociation, with a degree that increases with the temperature. The result obtained for the CE is consistent with this thermal behaviour. 
In fact,  the lower the CE, the more informationally stable  the state is, corresponding to a bigger dominance of the respective state in the system.  The probability of dissociation in the medium continuously  increases with the temperature. }}

For charmonium states, already in Fig. \ref{g67} one can realize a strong correlation between the CE and the temperature $T/T_c$ ratio. The higher the temperature, the higher the charmonium CE is. Besides, entropic Regge-like trajectories can be implemented in the context of AdS/QCD, that associates the charmonium CE with the temperature $T/T_c$ ratio.
In fact, interpolating the charmonium CE we find
 CE$\left(\frac{T}{T_c}\right) = -0.1008\frac{T^4}{T_c^4} +1.0103 \frac{T^3}{T_c^3}-3.7726\frac{T^2}{T_c^2}+6.7419 \frac{T}{T_c}-2.0590$, at the analyzed  temperature $T/T_c$ ratio range, within 0.08\%. It indicates a scaling law  between the CE and the temperature $T/T_c$ ratio, considering all the range $1\lesssim T_c\lesssim 3.5$.    \vspace*{0.5cm}
 \begin{figure}[H]
\begin{center}
\includegraphics[width=2.9in]{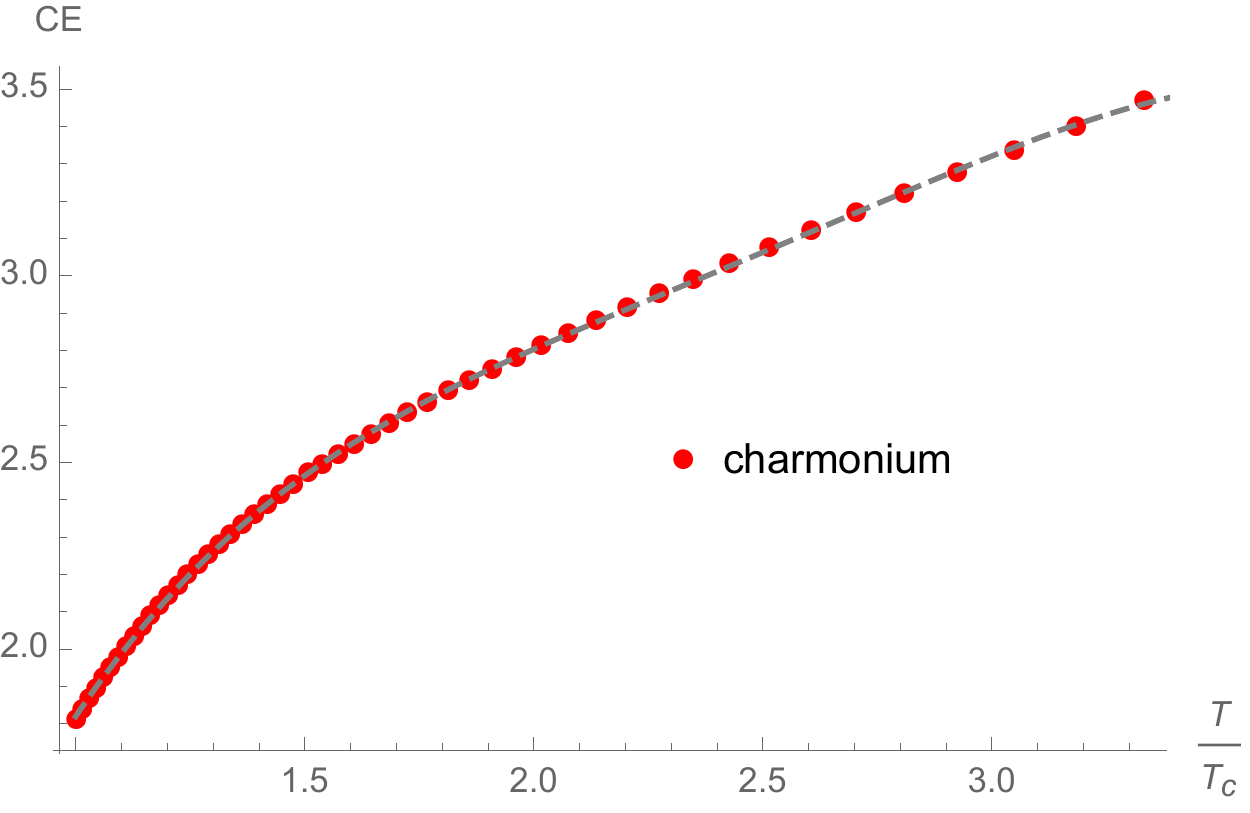}
\quad\quad
\caption{CE with respect to the temperature $T/T_c$ ratio, for the charmonium, and its quartic interpolation.}
 \label{g600}
\end{center}
\end{figure}

For the bottomonium 1$S$ state the result is different. Indeed, the CE decreases up to temperatures of the order of $ \sim 1.3\, T_c $ and then increases for higher temperatures. 
{\color{black}{Comparing with results from the analysis of the spectral function in Ref. \cite{Braga:2018zlu},   one should expect an increasing dissociation degree as the temperature increases.  Such a behaviour should be translated into a monotonic increase in the CE, although the bottomonium state survives at temperatures much higher than the charmonium. 
It is important to note, when comparing charmonium and bottomonium results, that the holographic model provides results with  different accuracy degrees for the two cases. From Tables \ref{table1} and \ref{table2}, one sees that for the charmonium 1$S$ state the  result for the mass differs just  5 \% from the experimental value. In contrast, for the bottomonium 1$S$ state the mass  of the model differs 27 \% from the experimental value. It is a very difficult task to find a simple holographic model that fits masses and decay constants for bottomonium. }}  It is possible that this problem with the fit of the value of the mass is 
leading to the failure of a consistent description of bottomonium dissociation, in terms of configurational entropy, for low temperatures $ T \lesssim 1.3 \,T_c $. 

The model represents a thermal medium, the plasma, that is present only for temperatures above the  critical one $T_c$. We considered, when plotting Fig. \ref{g67}, that the critical temperature  is $T_c = 191$ MeV \cite{Herzog:2006ra}, corresponding to the transition temperature, when the model is used for light flavor mesons. 
 
\begin{figure}[H]
\begin{center}
\includegraphics[width=2.9in]{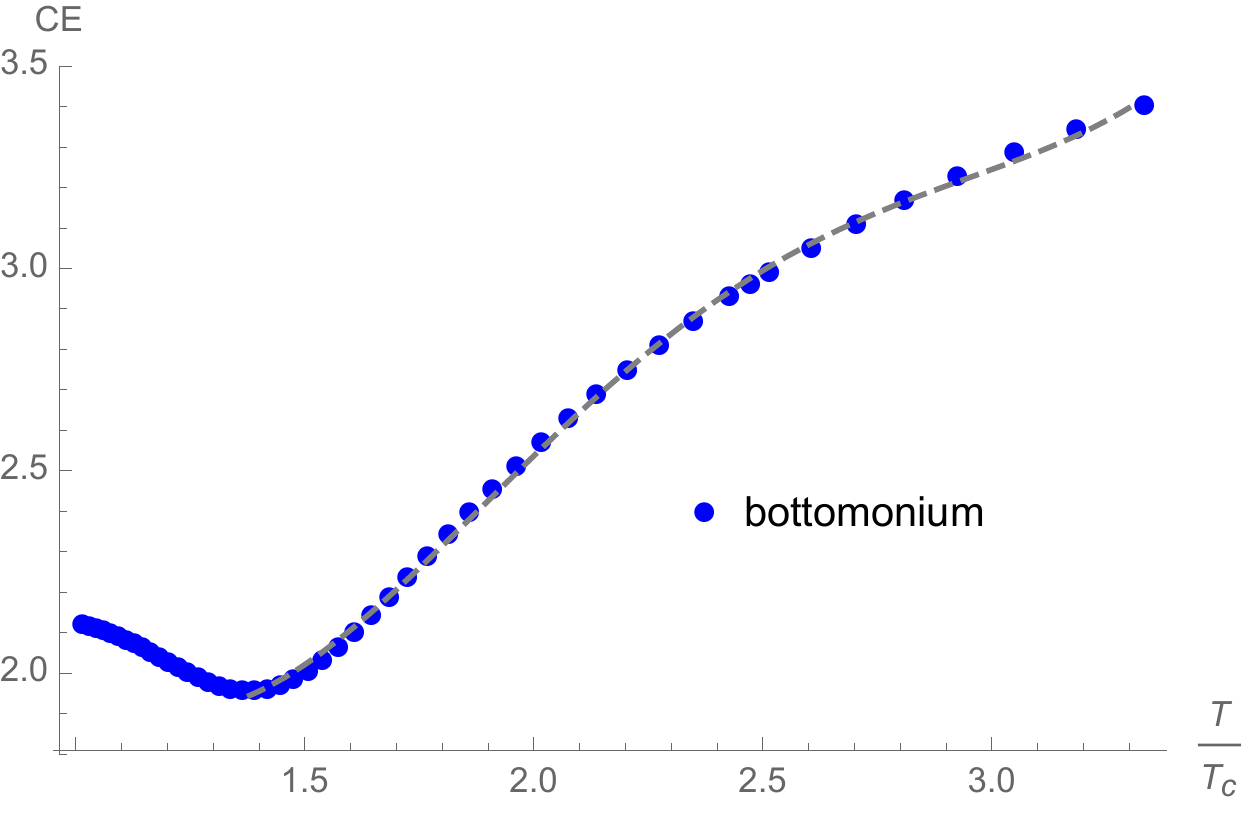}
\quad\quad
\caption{CE with respect to the temperature $T/T_c$ ratio, for the bottomonium, and its quartic interpolation.}
\label{g68}
\end{center}
\end{figure}

 A scaling law,  between the CE and the temperature $T/T_c$ ratio, also exists for the bottomonium state for temperatures  higher than $\sim 1.35 \,T_c$. The quartic polynomial CE$\left(\frac{T}{T_c}\right) = 0.2947\frac{T^4}{T_c^4} -2.8530 \frac{T^3}{T_c^3}+9.8565\frac{T^2}{T_c^2}-13.5219 \frac{T}{T_c}+8.2633$, at the range $1.35\lesssim T/T_c\lesssim 3.5$, fits the graphic for the bottomonium CE, within 0.29\%. 
This is plotted in Fig. \ref{g68}. 
   \vspace*{0.5cm}

\section{Conclusions}
 The dissociation of heavy vector mesons,  in the framework of holographic  AdS/QCD, was studied in the context of the configurational entropy setup. Charmonium and bottomonium states were analyzed from the point of view of  their quasinormal modes, at finite temperature. The quarkonia quasinormal modes are plotted in Figs. \ref{g40} -- \ref{g71}. The CE underlying quarkonia was analyzed both as a function of the temperature (Figs. \ref{g67} -- \ref{g68}) and of the radial excitation level of $S$-wave resonances (Tables \ref{table3} and \ref{table4}).  
 
 Important results make the quarkonia probability of dissociation to be distinct, with respect to the temperature. In fact, Figs. \ref{g67} --  \ref{g68} show the CE of bottomonium and charmonium as a function of the temperature $T/T_c$ ratio. The charmonium states  were shown to present their underlying  CE that increases monotonically with the temperature. In this case, the probability of dissociation in the medium increases with the temperature. For the bottomonium, Figs. \ref{g67} and \ref{g68} show a quite different result. First, the CE computed for the bottomonium has  a global minimum, corresponding to temperatures $T \sim 1.3\, T_c$. 
 Solely for higher temperatures, the CE increases monotonically. 
 One feasible explanation for this phenomenon 
 is that the  bottomonium   energy parameters are higher, compared to the charmonium ones. The AdS/QCD holographic model at finite temperature is assumed to represent a plasma when $T\gtrsim T_c$, at which  the deconfined medium is  formed.  The results for bottomonium indicate that only for $ T\gtrsim 1.35 T_c$ the model presents an appropriate description of the thermal deconfined medium. 
 It was possible to find for charmonium and bottomonium states, an interpolation of the CE as a function of the temperature $T/T_c$ ratio. Figs. \ref{g600} and \ref{g68} respectively show, for  the charmonium and the bottomonium,   scaling laws relating the CE to the temperature $T/T_c$ ratio. Our results indicate that the CE is a relevant tool to investigate stability of the heavy mesons against dissociation in the thermal medium.

\paragraph*{Acknowledgments:}   N.B. is partially supported by CNPq and L. F. Ferreira is supported CAPES; RdR~is grateful to Dr. A. Gon\c calves for fruitful discussions and to FAPESP (Grant No.  2017/18897-8) and to the National Council for Scientific and Technological Development  -- CNPq (Grant No. 303293/2015-2), for partial financial support.

\newpage

 \end{document}